\documentclass[a4paper,12pt]{article}
\usepackage[final]{graphics}
\usepackage{epsfig}
\usepackage{amsmath, amsthm, amssymb}
\usepackage{cite}
\numberwithin{equation}{section}
\begin{document}
\title{The brane universe as an Unruh observer}
\author{
David Jennings \footnote{D.Jennings@damtp.cam.ac.uk}, \\
\normalsize \em Department of Applied Mathematics and Theoretical Physics,\\
\normalsize \em CMS, University of Cambridge, Cambridge, CB3 0WA, UK. \\
}
\maketitle
\abstract{We make the observation that a brane universe accelerates through its bulk spacetime, and so may be interpreted as an Unruh observer. The bulk vacuum is perceived to be a thermal bath that heats matter fields on the brane. It is shown that, aside from being relevant in the early universe, an asymptotic temperature exists for the brane universe corresponding to late time thermal equilibrium with the bulk. In the simplest case two possible equilibrium points exist, one at the Gibbons-Hawking temperature for an asymptotic de Sitter universe embedded in an Anti-de Sitter bulk and another with a non-zero density on the brane universe. We calculate various limiting cases of Wightman functions in N-dimensional AdS spacetime and show explicitly that the Unruh effect only occurs for accelerations above the mass scale of the spacetime. The thermal excitations are found to be modified by both the curvature of the bulk and by its dimension. It is found that a scalar field can appear like a fermion in odd dimensions. We analyse the excitations in terms of vacuum fluctuations and back reactions and find that the Unruh effect stems solely from the vacuum fluctuations in even dimensions and from the back reactions in odd dimensions.}  
\newpage
\section{Introduction}
One might ask what the virtues of cosmology in higher dimensions are. In modern theoretical physics we have become accustomed to string theories where extra dimensions are required in order to have finite predictions. These unobserved dimensions are then often taken to be small, compact and outside our experimental grasp. \\
The idea of extra dimensions is not altogether new, but was originally considered by Kaluza and Klein\cite{kaluza} in a bid to geometrically combine gravity and electromagnetism. Now, with the emergence of $M$-theory, we have new models in which our universe is considered as a four dimensional hypersurface in a larger bulk spacetime. These brane scenarios stem from a string model suggested by Horava and Witten\cite{witten}, where the particles of the Standard Model are constrained to a hypersurface and remain four dimensional, while gravity is free to explore the full spacetime. \\
The motivations for studying these models include such things as a need to account for inflation and dark energy, and resolving the hierarchy problem\cite{randall1,randall2}. In many of these models the brane universe is embedded in a 5-D AdS spacetime. The expansion of the universe is now viewed as arising from the motion of the brane through the curved spacetime, where the scale factor $a(t)$ appears as a transverse coordinate describing the location of the brane. This motion of the brane through the bulk is determined by junction conditions\cite{israel} that relate the energy content of the brane to its dynamics. In general it is found that the brane accelerates through its bulk spacetime.\\
The concept of accelerated particle detectors in Minkowski spacetime was developed in the 1970s. In its simplest form, the particle detector is envisaged as a point particle with discrete internal energy levels, which is coupled to a quantum field.  
Unruh\cite{unruh} showed that such a detector moving with constant acceleration $a$, in Minkowski spacetime, perceives the vacuum state of the field to be a thermal reservoir with temperature $T=\frac{a}{2 \pi }$. This effect is not without its subtleties; for example a distinction must be made between particle content and energy transitions. It turns out that it is, in some sense, more meaningful to speak of energy excitations rather than particle content\cite{detectorclick, fulling}. The results for 4-dimensional Minkowski spacetime have been extended to curved spacetimes such as de Sitter spacetime where a generalised Gibbons-Hawking temperature\cite{gibbons_hawking} of $T=\sqrt{\frac{\Lambda}{3} +a^2}/2 \pi $ is found, with $\Lambda $ being the cosmological constant for the spacetime. \\
Quantum field theory in Anti-de Sitter spacetime possesses certain delicate features, such as its lack of global hyperbolicity\cite{isham}. Despite such difficulties it is found that an identical result to de Sitter spacetime holds, but this time the temperature switches off below the critical acceleration of $\sqrt{\frac{|\Lambda| }{3}}$. This result has been shown to hold in a general field theory framework where basic stability assumptions for the vacuum hold true\cite{passive_vacuum1,passive_vacuum2}. Explicit results for the transition rates of an Unruh detector in 4-D Anti-de Sitter spacetime have also been calculated\cite{levin}. \\
Since branes in general accelerate through their bulk spacetimes, we would expect that some form of the Unruh effect would arise, and cause energy excitations on the brane. We may then consider the simplest such model consisting of a single brane moving through an AdS bulk and allow it to couple to a bulk scalar field in its vacuum state. We aim to demonstrate explicitly the Unruh effect for N-dimensional AdS spacetime and calculate the energy excitations for a quantum system moving through such a spacetime.\\
We organise this paper as follows: In section 2 we establish some basic results concerning brane cosmologies. In section 3 we derive the necessary Wightman functions for a scalar field in N-dimensional AdS spacetime and analyse their limiting behaviours. In section 4 we calculate the detector response functions for basic stationary trajectories in AdS and demonstrate certain modifications to the Unruh effect that are dependent on the dimensionality of the spacetime and on the curvature of the spacetime. In sections 5 and 6 we analyse the excitations of an accelerated brane by calculating the separate contributions due to vacuum fluctuations and back reactions from the bulk spacetime. We show that the back reactions are altered in odd dimensions, while the vacuum fluctuations are altered in even dimensions and provide an acceleration dependent excitation of the matter fields on the brane. We observe that this effect would be relevant in the early universe when accelerations are large, but also in the late time universe where an asymptotic thermal equilibrium is achieved between the brane and the bulk. In section 7 we gather the main conclusions of this paper.  
\newpage
\section{Acceleration of the Brane}
In this section we review the dynamics of a single $Z_2$ -symmetric brane in a 5-D bulk spacetime\cite{carsten, langlois}.\\
The motion of the symmetric brane is determined by the Darmois-Israel junction conditions\footnote{This condition is not sufficient if the ${\mathbb Z}_2$ symmetry is relaxed, see for example\cite{bh} and \cite{graviton}.}, which state that the jump in its extrinsic curvature, $[K_{AB}]$, obeys the equation\cite{israel}
\begin{eqnarray}
[K_{AB}] = - \kappa ^2 ( T_{AB} -\frac{1}{3} h_{AB} T) 
\end{eqnarray}
where $T_{AB}$ is the energy momentum tensor for the brane and $h_{AB}$ is the induced metric. We now consider the tangential component of this extrinsic curvature, along its trajectory and define
\begin{eqnarray}
 K_{||}&=&u^Au^BK_{AB}.
\end{eqnarray}
It is a simple matter to show that 
\begin{eqnarray}
 K_{||}&=&-n_Ba^B
\end{eqnarray}
where $a^B = u^A\nabla_A u^B$, is the acceleration of the brane and $n^A$ is a unit vector field normal to the brane. Since we shall consider a co-dimension one brane, we deduce that
\begin{eqnarray}
 K_{||}&=&-a.
\end{eqnarray}
Using this, together with the junction condition and the mirror symmetry, we obtain that the acceleration of the brane is given by  
\begin{eqnarray}
 a&=&\frac{\kappa ^ 2}{6}( 2 \rho + 3 p - \sigma  )
\end{eqnarray}
where $ \sigma$ is the tension for the brane, $ \rho $ is its density and $p$ is the pressure. We see that in general the brane does not move along a geodesic.\\
A bulk spacetime that is consistent with demands of homogeneity and isotropy for the brane universe is 5-D AdS-Schwarzschild. We may obtain a modified FRW equation for a brane universe embedded in an AdS-Schwarzschild spacetime, whose metric is given by
\begin{eqnarray}
ds^2&=& -f(r)dt^2 + \frac{dr^2}{f(r)} + r^2 d \Omega ^2_3
\end{eqnarray}
where $d \Omega ^2_3$ is the metric on the maximally symmetric 3-space that possesses a scale factor $r(\tau )$, where $\tau $ is the proper time for the brane. If we take the universe to be flat, the function $f(r)$ is then given by 
\begin{eqnarray}
f(r)&=& k^2r^2 -\frac{C}{r^2}.
\end{eqnarray}
Using the spatial components of the junction conditions, we obtain
\begin{eqnarray}
H^2=\left (\frac{\dot{r}}{r}\right)^2&=& \frac{\kappa ^4}{36}\rho^2 + \frac{\kappa ^4 \sigma}{18}\rho + \frac{C}{r^4}+ \frac{\        \Lambda _4}{3}
\end{eqnarray}
This reproduces standard cosmology in the low density regime, but contains a quadratic density term and a dark radiation term that become relevant in the very early universe. The effective cosmological constant for the brane universe is given in terms of the brane tension and AdS mass parameter as 
\begin{eqnarray}
\Lambda _4=3\left (\frac{\kappa ^4 \sigma^2 }{36} -k^2 \right ).
\end{eqnarray}\\
Since the brane accelerates through its bulk spacetime we might expect it to experience some form of Unruh radiation which would produce energy excitations of the fields on the brane. 
It can be shown\cite{birrell_davies} that, to lowest order in the coupling constant $c$, the probability amplitude for transitions from the ground state $|E_0>$ of a particle detector coupled to a scalar field in its vacuum state $|0>$ is given by
\begin{eqnarray}
c^2 \sum_E |<E|M|E_0>|^2  {\cal F}(E) 
\end{eqnarray} 
where $M$ is the detector observable that is coupled to the field, and we define the detector response function ${\cal F }(E)$ and the Wightman function ${\cal G}^+(x,x')$ as 
\begin{eqnarray}
{\cal F}(E) &=& \int dt dt'{\cal G}^+(x(t),x(t'))e^{-i(E-E_0)(t-t')}\nonumber\\
{\cal G}^+(x,x') &=& <0|\phi (x) \phi (x') |0>.
\end{eqnarray}
We see that if the detector response function does not vanish then excitations will be measured by the detector. In many cases the detector is placed on a stationary trajectory for which the Wightman function becomes a function of $\Delta \tau =\tau - \tau '$, where $\tau $ is the proper time along the path. In this case the response function is ill-defined and it makes physical sense to consider instead the response function per unit time
\begin{eqnarray}
 \frac{ {\cal F}(E)}{T}&=& \int^\infty_{-\infty} d \Delta \tau e^{-i(E-E_0)\Delta \tau } {\cal G}^+(\Delta \tau ) .
\end{eqnarray}  
In the next section we shall derive the Wightman functions for a scalar field in Anti-de Sitter spacetime that are necessary to discuss any Unruh effect that a brane, coupled to such a field, would experience.  
\section{Scalar field in N-dimensional AdS}
In recent times an extensive body of work has grown up around Anti-de Sitter spacetime, and it is this spacetime that often provides the setting for brane models of cosmology. It is necessary to derive some basic results that are needed in order to deal with a brane coupled to any bulk fields. It is no great inconvenience to talk in generalities, and so we may consider an N dimensional AdS spacetime with constant Ricci scalar $R=-N(N-1)k^2$. The metric for this spacetime may be expressed in Poincar\'{e} coordinates as
\begin{eqnarray}
 ds^2&=&e^{-\frac{2y}{k}}( dt^2 - dx_1^2 - dx_2^2- \cdots -dx_{N-2}^2) -dy^2
\end{eqnarray}
which is made conformally flat by the change of variables
\begin{eqnarray}
z&=&\int e^{\frac{y}{k}}dy
\end{eqnarray}
to produce
\begin{eqnarray}
ds^2&=&\frac{1}{k^2 z^2}( dt^2 - dx_1^2 - dx_2^2- \cdots -dx_{N-2}^2 -dz^2).
\end{eqnarray}
The equation of motion of a bulk scalar field of mass $m$, coupled to gravity in this spacetime, is given by
\begin{eqnarray}
(\nabla^\mu \nabla_\mu +m^2 +\zeta R) \phi(x) &=&0 
\end{eqnarray}
where $\zeta$ is its coupling to gravity.\\
Making use of the planar symmetry, it can be seen that modes exist of the form\cite{ads_wightman}
\begin{eqnarray}
 \phi(x)&=& \phi_{\textbf{p}}(x^i)\beta_n(y)\nonumber \\
 \phi_{\textbf{p}}(x^i)&=& \frac{e^{-i\eta_{ij}p^ix^j}}{\sqrt{2\omega(\textbf{p}) (2\pi)^N-2}}
\end{eqnarray}
where $\eta_{ij}$ is the metric for an $N-1$ dimensional Minkowski spacetime and 
\begin{eqnarray}
 p^i&=&(\omega(\textbf{p}), \textbf{p}) \nonumber\\
\omega(\textbf{p}) &=& \sqrt{|\textbf{p}|^2 + m_n^2}
\end{eqnarray}
The constants $m_n$ are obtained from a separation of variables in the equation of motion and are determined once boundary conditions are specified.\\
The functions $\beta_n(y)$, which describe the field transverse to the Minkowski planes, are given by
\begin{eqnarray}
 \beta_n(y)&=&c_n e^{(N-1)\frac{ky}{2}}(J_\nu(m_nz) + b_\nu Y_\nu(m_nz))
\end{eqnarray}
where $J_\nu(m_nz)$ and  $ Y_\nu(m_nz)$ are Bessel and Neumann functions respectfully.\\
The parameter $\nu $ is of strong importance and behaves like an effective scale for the field. It is given by
\begin{eqnarray}
 \nu &=& \sqrt{\frac{(N-1)^2}{4} -N(N-1)\zeta +\frac{m^2}{k^2}} .
\end{eqnarray}
\subsection{Wightman functions}
The coupling of the scalar field to an observer moving in this spacetime is governed by a two-point function of the field. More specifically, when the field is in a vacuum state the Wightman function provides the necessary information. The Wightman function for a scalar field in a vacuum state $|0>$, is defined as
\begin{eqnarray}
 {\cal G}^+(x,x')&=&<0|\phi(x)\phi(x')|0>.
\end{eqnarray}
For the case of AdS spacetime, it is possible to write this as\cite{ads_wightman}
\begin{eqnarray}
 {\cal G}^+_{AdS}(x,x')&=&k^{N-2} (zz')^{\frac{N-1}{2}}\int_0^\infty dm \left [m{\cal G}^+_{{\mathbb M}_{N-1}}(x,x';m)J_\nu(mz)J_\nu(mz')\right ] \nonumber
\end{eqnarray}
where ${\cal G}^+_{{\mathbb M}_{N-1}}(x,x';m)$ is the Wightman function for a scalar field, of mass $m$, in an $(N-1)$-dimensional Minkowski spacetime. The explicit form of this lower dimensional Wightman function is
\begin{eqnarray}
 {\cal G}^+_{{\mathbb M}_{N-1}}(x,x';m)&=& \frac{m^{\alpha-\frac{1}{2}}K_{\alpha -\frac{1}{2}} (m\sqrt{(\textbf{x}-\textbf{x}')^2 - (t-t'-i \epsilon )^2})}{(2\pi)^{\alpha +\frac{1}{2}}((\textbf{x}-\textbf{x}')^2 - (t-t'-i \epsilon )^2)^{\frac{\alpha}{2} -\frac{1}{4}}}\nonumber
\end{eqnarray}
where $K_\nu(x)$ is a modified Bessel function, $\epsilon $ is, as usual, a small positive constant and we have introduced for convenience $\alpha =\frac{N}{2}-1$.\\
We then find that the Wightman function for a massive scalar field in $N$-dimensional AdS spacetime is then given by
\begin{eqnarray}\label{wightman}
  {\cal G}^+_{AdS_{N}}(x,x')&=& \frac{k^{2 \alpha }e^{-\alpha \pi i}}{(2\pi)^{\alpha +1}}\frac{1}{(v^2-1)^{\frac{\alpha} {2}}}Q^\alpha_{\nu -\frac{1}{2}} (v)
\end{eqnarray}
where
\begin{eqnarray}\label{v}
 v&=&\frac{z^2 + z'^2 + (\textbf{x} -\textbf{x}')^2 -(t-t'-i \epsilon )^2}{2zz'} 
\end{eqnarray}
and $Q^\alpha_\beta (x)$ is the associated Legendre function of the second kind.
\subsection{Limits on mass and curvature}
The form of the Wightman function for a massive scalar field is relatively intractable for our present needs, but we shall see that the corresponding \emph{massless} case takes on a much more flexible form. For completeness, we shall compute the two different limiting routes by which we may pass to a massless field in $N$ dimensional Minkowski spacetime.
\subsubsection{Massless scalar field in Minkowski spacetime}
To calculate the Wightman function for a massless scalar field in the Minkowski spacetime we may first reduce the curvature of the spacetime to zero and then allow the mass of the field to vanish.\\
It is shown in \cite{ads_wightman} that (\ref{wightman}) tends to ${\cal G}^+_{{\mathbb M}_N}$ in the limit $k\rightarrow 0$. Namely,
\begin{eqnarray}
\lim_{k\rightarrow 0} {\cal G}^+_{AdS_N}(x,x') &=&{\cal G}^+_{{\mathbb M}_N}(x,x')= \frac{m^\alpha}{(2\pi)^{\alpha +1}w^\alpha }K_\alpha (mw)  \nonumber
\end{eqnarray}
where $w^2 = (y-y')^2 +(\textbf{x} +\textbf{x}')^2 -(t-t'-i \epsilon )^2$ and $K_\alpha(x)$ is once again the modified Bessel function.\\
To compute the massless limit of this we observe that for small $x$
\begin{eqnarray}
 K_\alpha (x)&\sim& \frac{1}{2} \Gamma(\alpha )\left ( \frac{2}{x}\right )^\alpha  
\end{eqnarray}
from which we find
\begin{eqnarray}
\lim_{m \rightarrow 0} {\cal G}^+_{ {\mathbb M}_N}(x,x') &=& \frac{\Gamma (\alpha )}{4 \pi^{\alpha +1} w^{2 \alpha }}.
\end{eqnarray}
\subsubsection{Massless scalar field in AdS spacetime}
We now calculate the Wightman function for a massless scalar field in AdS spacetime, which will be of central importance to this paper.
To compute this function we need a well behaved expansion for $Q^\alpha _\beta (x)$ in the limit of vanishing mass. One such expansion is
\begin{eqnarray}
 Q^\alpha _\beta (x)&=&\frac{e^{\alpha \pi i}}{2^{\beta +1}}\frac{\Gamma( \alpha + \beta + 1)}{\Gamma(\beta +1)}(x^2 -1)^{\frac{\alpha}{2}} \int^1_{-1}\frac{(1-t^2)^\beta}{ (x -t)^{\alpha + \beta + 1}}dt \nonumber.
\end{eqnarray}
With this we find that
\begin{eqnarray}
 \lim_{m \rightarrow 0} {\cal G}^+_{AdS_N}(x,x')&=& \frac{k^{2 \alpha }}{(2 \pi )^{\alpha +1}2^{\nu_0 +\frac{1}{2}}}\frac{\Gamma( \alpha + \nu_0 +\frac{1}{2})}{\Gamma (\nu_0 +\frac{1}{2})} \int^1_{-1}\frac{(1-t^2)^{\nu_0 -\frac{1}{2}}}{(v-t)^{\alpha + \nu_0 +\frac{1}{2}}}dt \nonumber
\end{eqnarray}
where $\nu_0 = \sqrt{\frac{(N-1)^2}{4} -N(N-1)\zeta}$. 
This is rather messy. We now make the further assumption that the massless scalar field is conformally coupled to gravity, and consequently $\nu_0 =\frac{1}{2}$. We conclude that
\begin{eqnarray}\label{G}
  \lim_{m \rightarrow 0} {\cal G}^+_{AdS_N}(x,x')&=&{\cal G}^+_{AdS_N,m=0}(x,x')\nonumber \\ 
&=& \frac{k^{2 \alpha } \Gamma (\alpha)}{2(2\pi)^{\alpha +1}} \left ( \frac{1}{(v-1)^ \alpha } - \frac{1}{(v+1)^ \alpha }  \right ).
\end{eqnarray}\\
This result behaves correctly as we shrink the AdS curvature to zero. For small $k$, $v \sim 1 +\frac{k^2w^2}{2}$ and consequently
\begin{eqnarray}
{\cal G}^+_{AdS_N,m=0}(x,x') &\sim&  \frac{k^{2 \alpha } \Gamma (\alpha)}{2(2\pi)^{\alpha +1}} \left ( \frac{2^\alpha }{k^{2 \alpha}w^{2 \alpha } } - \frac{1}{(2 + \frac{k^2w^2}{2})^ \alpha }\right )\nonumber
\end{eqnarray}
Taking the limit as the AdS curvature tends to zero gives
\begin{eqnarray}
 \lim_{k\rightarrow 0}  {\cal G}^+_{AdS_N,m=0}(x,x')&=&\frac{\Gamma (\alpha )}{4 \pi^{\alpha +1} w^{2 \alpha }}
\end{eqnarray}
in agreement with our previous result on the massless limit of the Minkowski Wightman function.\\
To sum up, we have arrived at the following limits for Wightman functions in $N$-dimensional AdS spacetime:\\

\begin{tabular}{|rcccl|}
\hline
&&&&\\
& & \tiny $ (k\rightarrow 0) $ \normalsize & & \\
${\cal G}^+_{AdS_N} \propto \frac{1}{(v^2 -1)^{\frac{\alpha}{2}}}Q^\alpha _{\nu -\frac{1}{2}}(v)$& &$\longrightarrow$ & & ${\cal G}^+_{{\mathbb M}_N} \propto \frac{m^\alpha }{w^\alpha }K_\alpha (mw)$ \\
$\Large \downarrow \normalsize $ &\tiny $(m \rightarrow 0)$  \normalsize & &  \tiny $(m \rightarrow 0)$ \normalsize &$ \Large \downarrow \normalsize $ \\
& & \tiny $ (k\rightarrow 0) $ \normalsize & & \\
${\cal G}^+_{AdS_N,m=0} \propto \frac{1}{(v-1)^\alpha} - \frac{1}{(v+1)^\alpha}$ && $\longrightarrow$ & & ${\cal G}^+_{{\mathbb M}_N,m=0} \propto \frac{1 }{w^{2\alpha} }$ \\
&&&&\\
\hline
\end{tabular}\\

with $\alpha = \frac{N}{2}-1$ specifying the dimensionality of the system, $\nu $ an effective mass for the scalar field and separation of spacetime points given by $v$ and $w$.
\newpage
\section{Unruh effect for AdS branes} 
In the previous section we derived results relating to the bulk scalar field. In this section we consider a brane, coupled to this bulk scalar field, that is moving along certain stationary trajectories in the AdS spacetime. The most basic stationary trajectories are the timelike geodesic, and the constant acceleration trajectories. We expect the response function per unit time to vanish along geodesics, and consequently the inertial brane experiences no excitation from the bulk vacuum. We shall also find that excitations only occur above a certain critical acceleration. For simplicity we assume that the scalar field is conformally coupled to gravity. We shall find that it behaves like a fermion in odd dimensions and a boson in even dimensions. The relevant Wightman function for such a field is given by (\ref{G}) and for simplicity we take the ground state energy $E_0$ to be zero.
\subsection{Geodesics}
We work with the conformally flat metric
\begin{eqnarray}
ds^2&=&\frac{1}{k^2z^2}( dt^2 - dx_1^2 - dx_2^2- \cdots -dx_{N-2}^2 -dz^2)
\end{eqnarray}
and restrict motion to the $z-t$ plane. The timelike geodesics may be written as
\begin{eqnarray}
 \gamma k t &=& \tan (k \tau + A )\nonumber\\
 \gamma k z &=& \sec (k \tau+ A)
\end{eqnarray}
where $\gamma$ and $A$ are constants and $\tau$ is the proper time. \\
We wish to consider ${\cal G}^+(x(\tau),x(\tau '))$ and expect it to be purely a function of $\Delta \tau = \tau -\tau '$.
To this end we must calculate $v(\tau,\tau ')$, given by (\ref{v}), along the trajectory. It is easily seen that
\begin{eqnarray}
 v(\tau,\tau ')&=&\cos(k \Delta \tau - i \epsilon)
\end{eqnarray}
and substitution into (\ref{G}) gives that
\begin{eqnarray}
 {\cal G}^+(\Delta \tau)&=& B\left [ \frac{1}{(\cos (k \Delta \tau -i \epsilon )  -1)^{\frac{N}{2}-1} }- \frac{1}{(\cos (k \Delta \tau -i \epsilon )  +1)^{\frac{N}{2} -1} } \right ]\nonumber
\end{eqnarray}
where
\begin{eqnarray}
 B&=& \frac{k^{N-2 } \Gamma (\frac{N}{2}-1)}{2(2\pi)^{\frac{N}{2}}}.
\end{eqnarray}
All the poles of this function lie in the upper half plane, at the points $\frac{n \pi}{k} + i \epsilon $, $n \in {\mathbb Z}$. Since $E>0$ the integral
\begin{eqnarray}
 \int^\infty_{-\infty} e^{-iE\Delta \tau} {\cal G}^+(\Delta \tau) d \Delta \tau
\end{eqnarray}
may be evaluated by closing a contour in the lower half plane. There is a complete absence of poles in this region, and so we deduce that the integral vanishes. No excitations are observed, as is expected for a geodesic path.
\subsection{Trajectories of constant acceleration}\label{wight}
Although no excitations are observed for an inertial brane, it is expected that some form of the Unruh effect will occur for accelerated trajectories through the curved spacetime. The simplest of these trajectories are those where the acceleration remains constant throughout.\\
By considering planar intersections with AdS in a flat embedding space it is possible to obtain all constant acceleration curves\cite{fluxoid}. These curves are effectively conic sections of the AdS spacetime. There are three classes of constant acceleration timelike curves: elliptic, parabolic and hyperbolic. Elliptic trajectories are those curves for which the acceleration, $a$, satisfies $a^2 < k^2$, parabolic curves are those that satisfy $a^2=k^2$, while hyperbolic trajectories satisfy $a^2 >k^2$. The acceleration is then said to be subcritical, critical or supercritical respectively.\\
\subsubsection{Subcritical accelerations}
We may describe trajectories with accelerations below $k$ in the Poincar\'{e} coordinate system by
\begin{eqnarray}
 t( \tau )&=& z_0 k \sinh \gamma (\tau ) \nonumber\\ 
 z( \tau )&=& z_0 k \cosh \gamma (\tau )+z_0a \nonumber\\ 
\end{eqnarray}
with
\begin{eqnarray}
 \cosh \gamma ( \tau )&=& \frac{a \cos( \sqrt{k^2 -a^2} \tau ) - k}{a- k \cos ( \sqrt{k^2 -a^2} \tau )}.
\end{eqnarray}
Evaluating at two points along a given trajectory we obtain that
\begin{eqnarray}
 v( \tau , \tau ')&=&\frac{k^2}{k^2 -a^2} \cos( \sqrt{k^2 -a^2} \Delta \tau - i \epsilon ) 
\end{eqnarray}  
which gives that 
\begin{eqnarray}
 {\cal G}^+(\Delta \tau)&=&  \left ( \frac{B}{(\frac{k^2}{k^2 -a^2} \cos( \sqrt{k^2 -a^2} \tau  - i \epsilon ) -1)^{\frac{N}{2}-1}} -\frac{B}{(\frac{k^2}{k^2 -a^2} \cos( \sqrt{k^2 -a^2} \tau  - i \epsilon ) +1)^{\frac{N}{2}-1}} \right ).\nonumber
\end{eqnarray}
No poles exist in the lower half plane, and no excitations from the ground state are detected for subcritical accelerations.
\subsubsection{Critical accelerations}
Trajectories with critical accelerations may be described as
\begin{eqnarray}
 z&=&z_0\nonumber \\
t &=& kz_0\tau  .
\end{eqnarray}
For two points along such a trajectory with have
\begin{eqnarray}
 v(\tau,\tau')&=&1- \frac{k^2}{2} ( \Delta \tau -i \epsilon )^2 
\end{eqnarray}
and so
\begin{eqnarray}
 {\cal G}^+(\Delta \tau)&=&B\left ( \frac{2^{N-2}}{k^{N-2} (\Delta \tau - i \epsilon )^{N-2}} - \frac{1}{(2 - \frac{k^2}{2}(\Delta \tau -i \epsilon )^2)^{\frac{N}{2}-1}}  \right )\nonumber.
\end{eqnarray}
It is clear that once again all singularities lie in the upper half plane, and so there will be no excitations detected. We also note that the first term is identical to the result obtained for a \textit{geodesic} trajectory in \textit{Minkowski} spacetime. 
\subsubsection{Supercritical accelerations}
The supercritical trajectories are given by
\begin{eqnarray}
 t&=& \frac{az_0}{\omega }e^{\omega \tau}\nonumber \\
 z&=& z_0e^{\omega \tau}
\end{eqnarray}
with $\tau$ once again being the proper time along the trajectory and $\omega =\sqrt{a^2 -k^2}$.\\
By considering two points along a given trajectory we obtain that
\begin{eqnarray}
 v(\tau, \tau ')&=&\frac{a^2}{\omega ^2} -\frac{ k^2}{\omega ^2} \cosh(\omega \Delta \tau - i \epsilon ).
\end{eqnarray}
Substituting this into (\ref{G}) gives us that the Wightman function along the trajectory is
\begin{eqnarray}
{\cal G}^+(\Delta \tau)&=& \frac{\omega ^{N-2} \Gamma (\frac{N}{2}-1)}{(4\pi)^{\frac{N}{2}}} [ \frac{1}{i^{N-2}(\sinh ^{N-2}(\frac{\omega \Delta \tau}{2} -i \epsilon)}\nonumber \\
& & - \frac{1}{(\sinh(A + (\frac{\omega \Delta \tau}{2} -i \epsilon) ))^{\frac{N}{2}-1}  (\sinh(A - (\frac{\omega \Delta \tau}{2} -i \epsilon) ))^{\frac{N}{2}-1}}  ]\nonumber
\end{eqnarray}
with $\sinh A = \frac{\omega}{ k}$.\\
The second term does not feature in the limit to Minkowski spacetime, and is purely from the non-zero curvature that the AdS spacetime possesses.\\
We now proceed to calculate the detector response function per unit time for this path. We divide the Wightman function into two terms, ${\cal G}^+(\Delta \tau) = {\cal G}^+_1(\Delta \tau)- {\cal G}^+_2(\Delta \tau)$, and calculate the contribution from each term separately. 
\begin{eqnarray}
{\cal G}^+_1(\Delta \tau)&=& \frac{\omega ^{N-2} \Gamma (\frac{N}{2}-1)}{(4\pi)^{\frac{N}{2}}} \left [ \frac{1}{i^{N-2}(\sinh ^{N-2 }(\frac{\omega \Delta \tau}{2} -i \epsilon)} \right ] \nonumber\\
{\cal G}^+_2(\Delta \tau)&=& \frac{\omega ^{N-2} \Gamma (\frac{N}{2})}{(4\pi)^{\frac{N}{2}}} \left [\frac{1}{(\sinh(A + (\frac{\omega \Delta \tau}{2} -i \epsilon) ))^{\frac{N}{2}-1}(\sinh(A - (\frac{\omega \Delta \tau}{2} -i \epsilon) ))^{\frac{N}{2}-1} }  \right ] \nonumber.
\end{eqnarray}
 We first evaluate the contribution due to ${\cal G}^+_1$.  The integrand of 
\begin{eqnarray}
 \int^\infty_{-\infty} e^{-iE \Delta \tau} {\cal G}^+_1(\Delta \tau) d \Delta \tau = \frac{\omega ^{N-2} \Gamma (\frac{N}{2}-1)}{(4\pi)^{\frac{N}{2}}}\int^\infty_{-\infty} \frac{e^{-iE \Delta \tau}}{\sinh ^{N-2} (\frac{\omega \Delta \tau}{2} -i \epsilon )} d \Delta \tau \nonumber
\end{eqnarray}
has poles of order $(N-2)$ all along the imaginary axis at the points $\Delta \tau = i( \epsilon -\frac{2m\pi}{ \omega} )$, with $m \in {\mathbb Z}$. The residues may be calculated by standard methods and are found to depend on a real polynomial $f_{N-3}(\frac{E}{\omega })$, of order $N-3$. Explicit forms for these polynomials are given in the appendix. \\
Closing a contour in the lower half plane allows us to obtain the result
\begin{eqnarray}
 \int^\infty_{-\infty}e^{-iE \Delta  \tau}{\cal G}^+_1(\Delta \tau) d \Delta \tau&=& \frac{\Gamma(\frac{N}{2}-1) \omega ^{N-3} f_{N-3}(\frac{E}{ \omega} )}{(4 \pi )^{\frac{N}{2}}}\frac{1}{e^{2 \pi \frac{E}{ \omega} } - (-1)^N}\nonumber
\end{eqnarray}
and so we obtain a thermal distribution which is either \emph{fermionic} or \emph{bosonic}, depending on the dimensionality of the spacetime. The origin of the fermionic statistics may be traced back to the Wightman function being \emph{anti-periodic} in odd dimensions.\\
The contribution due to ${\cal G}^+_2$ may be dealt with in a similar way. In this case the function is slightly more complicated. The function
\begin{eqnarray}
{\cal G}^+_2(\Delta \tau)&=& \frac{\omega ^{N-2} \Gamma (\frac{N}{2}-1)}{(4\pi)^{\frac{N}{2}}} \left [\frac{1}{(\sinh(A + (\frac{\omega \Delta \tau}{2} -i \epsilon) ))^{\frac{N}{2}-1} (\sinh(A - (\frac{\omega \Delta \tau}{2} -i \epsilon) ))^{\frac{N}{2}-1} }  \right ] \nonumber
\end{eqnarray}
has poles along two lines in the complex plane. Specifically, it is singular at 
\begin{eqnarray}
\Delta \tau_{m, \pm} = i \epsilon \pm \frac{2A}{\omega } -\frac{2m\pi i}{\omega }.
\end{eqnarray}
For even dimensional spacetimes these poles are isolated and the residue theorem may be employed. Calculation of the residue at $\Delta \tau_{m,+}$ plus the residue at $\Delta \tau_{m,-}$ gives 
\begin{eqnarray}
 \frac{\Gamma (\frac{N}{2}-1)ie^{-2m\pi \frac{E}{\omega} }}{2\pi(4\pi)^{\frac{N}{2}}} k^{N-2}( q_1(E,a,k) \sin \theta + q_2(E,a,k) \cos \theta )
\end{eqnarray}
where
\begin{eqnarray}
 \theta &=& \frac{2E}{\omega }\sinh^{-1} (\frac{ \omega }{k})
\end{eqnarray}
and $q_1$ and $_2$ are functions analogous to $f_{N-3}$. The full contribution from ${\cal G}^+_2$ is obtained by summing over those poles in the lower half plane 
\begin{eqnarray}
  \int^\infty_{-\infty} d \Delta \tau {\cal G}^+_2(\Delta \tau)&=& \frac{k^{N-2} \Gamma(\frac{N}{2}-1)  }{(4 \pi )^{\frac{N}{2}}}\frac{q_1(E,a,k) \sin \theta + q_2(E,a,k) \cos \theta}{e^{2 \pi \frac{E}{ \omega} } - 1}. \nonumber
\end{eqnarray}
Explicit forms for the functions $q_1$ and $q_2$ are also given in the appendix. It is clear that the contribution due to ${\cal G}^+_2$ does indeed vanish as the curvature of the AdS spacetime shrinks to zero.\\
 For odd dimensions branch points exist, but we may deduce the statistical denominator of our results from the anti-periodicity of ${\cal G}^+_2$,
\begin{eqnarray}
 G_2^+(\Delta \tau + \frac{2 \pi}{\omega} )&=& (-1)^N  G_2^+(\Delta \tau )
\end{eqnarray}
which implies that in general 
\begin{eqnarray}
 \int^\infty_{-\infty} d \Delta \tau {\cal G}^+_2(\Delta \tau)&=&\frac{\Gamma(\frac{N}{2}-1) k^{N-2}}{(4 \pi )^{\frac{N}{2}}}\frac{g(E,a,k)}{e^{2 \pi \frac{E}{ \omega} } - (-1)^N}
\end{eqnarray}
where $g(E,a,k)$ is determined by a contour integral around singular points which may or may not be branch points.\\
By adding together the two contributions we obtain the result
\begin{eqnarray}
\int^\infty_{-\infty} d \Delta \tau {\cal G}^+(\Delta \tau)&=&\frac{F}{e^{2 \pi \frac{E}{ \omega} } - (-1)^N}
\end{eqnarray}
where 
\begin{eqnarray}
 F&=& \frac{\Gamma (\frac{N}{2}-1)}{(4 \pi )^{\frac{N}{2}}} (\omega ^{N-3}f_{N-3}(\frac{E}{\omega }) - k^{N-2}g(E,a,k)).
\end{eqnarray}
We conclude that an observer moving along a hyperbolic trajectory in AdS spacetime, with an acceleration exceeding $k$, will perceive the vacuum state of the scalar field to be a modified thermal one, with a temperature
\begin{eqnarray}
 T&=&\frac{\sqrt{a^2 - k^2}}{2 \pi } .
\end{eqnarray}
The modified excitations that the coupling produces depend on the dimensionality of the bulk spacetime and its curvature. The statistics of the field, as perceived by the observer, are fermionic in odd dimensions and bosonic in even dimensions.
\newpage
\section{Vacuum Fluctuations and Back Reactions}
To analyse in more detail the excitations due to the non-inertial motion of the universe through the AdS spacetime, we present a model of brane matter coupled to a scalar field in the bulk. We follow \cite{accelerated_atom} and proceed to separate these excitations into those that stem primarily from fluctuations in the bulk vacuum, and those that are due to the disturbance of the bulk scalar field resulting from the matter coupling.\\
The model consists of a quantum system with no internal spatial variations, discrete multiple energy levels $\{ E_i \}$ and that is coupled to a scalar field $\phi (x)$ by a term $c M(\tau) \phi(x(\tau))$, where $\tau$ is the proper time along the trajectory of the multilevel system. This discrete system represents the accelerated brane. The Hamiltonian for the whole model is then
\begin{eqnarray}
 H_{\mbox{\tiny total \normalsize}}&=& H(\tau) + \int d^3k \frac{dt}{d \tau} a^\dagger(\textbf{k},t)a(\textbf{k},t)\omega( \mathbf{k}) + cM(\tau) \phi (x(\tau)) \nonumber
\end{eqnarray}
with $H(\tau)$ the Hamiltonian for the discrete system and $\{a(\mathbf{k},t)\}$ a mode decomposition for the scalar field.\\
We now divide each physical variable into a \emph{free} part and a \emph{source} part. The free part is present as $c\rightarrow 0$, while the source part is first order in $c$ - its presence arising from the exchange of energy via the monopole coupling
\begin{eqnarray}
 H(\tau)&=& H_0(\tau) + H_s(\tau)\nonumber\\
 M(\tau)&=& M_0(\tau) + M_s(\tau)\nonumber\\
 \phi(\tau)&=& \phi_0(\tau) + \phi_s(\tau).
\end{eqnarray}
The source part of an arbitrary brane operator, $X$, is then given, to first order in $c$, by
\begin{eqnarray}
X_s(\tau) &=& ic \int^\tau_{\tau_0} d \tau ' \phi_0(x(\tau '))[M_0(\tau '),X_0(\tau )] 
\end{eqnarray}
while the source part of an arbitrary field operator, $Y$, is
\begin{eqnarray}
Y_s(x(\tau)) &=& ic \int^\tau_{\tau_0} d \tau 'M_0(\tau ') [\phi_0(x(\tau ')),Y_0(\tau )] .
\end{eqnarray}
For a given observable, X, of the discrete system we may consider the contribution to its time evolution due to the scalar field $\phi $. Explicitly we have
\begin{eqnarray}
\left ( \frac{dX(\tau)}{d \tau} \right)_\phi&=& i c \phi(x(\tau))[M(\tau),X(\tau)]  .
\end{eqnarray}
A decomposition of this variation into a part resulting from the free field term and another resulting from the source field term is now possible. These contributions will be the vacuum fluctuation and back reaction respectively. A subtlety arises when we attempt to perform this decomposition. To arrive at a physically unambiguous and well-defined notion of vacuum fluctuations as separate from back reaction it is necessary to use a symmetric form of the operator ordering, namely
\begin{eqnarray}
\left ( \frac{dX(\tau)}{d \tau} \right)_\phi &=& i \frac{c}{2} \left ( \phi(x(\tau))[M(\tau),X(\tau)] + [M(\tau),X(\tau)] \phi(x(\tau))\right ).\nonumber
\end{eqnarray}
From the point of view of the total variation, this is a trivial reordering, since operators for the scalar field and operators for the brane commute, but has significance when we restrict our attentions to the individual terms.\\
The vacuum fluctuations and back reaction are then given, in a well-defined manner, by the expressions
\begin{eqnarray}
\left ( \frac{dX(\tau)}{d \tau} \right)_{\mbox{\tiny vac \normalsize}}&=& i \frac{c}{2} \left ( \phi_0(x(\tau))[M(\tau),X(\tau)] + [M(\tau),X(\tau)] \phi_0(x(\tau))\right )\nonumber\\
\left ( \frac{dX(\tau)}{d \tau} \right)_{\mbox{\tiny back \normalsize}}&=& i \frac{c}{2} \left ( \phi_s(x(\tau))[M(\tau),X(\tau)] + [M(\tau),X(\tau)] \phi_s(x(\tau))\right ).\nonumber
\end{eqnarray}
Expressing everything, to order $c^2$, in terms of all the free fields, we find that
\begin{eqnarray}
\left ( \frac{dX(\tau)}{d \tau} \right)_{\mbox{\tiny vac \normalsize}}&=& i \frac{c}{2} ( \phi_0(x(\tau))[M_0(\tau),X_0(\tau)] + [M_0(\tau),X_0(\tau)] \phi_0(x(\tau)))\nonumber\\
&-&\frac{c^2}{2}\int^\tau_{\tau_0}d\tau '\{\phi_0(x(\tau )) , \phi_0(x(\tau )) \}[M_0(\tau'),[M_0(\tau),X_0(\tau)]] \nonumber\\
\left ( \frac{dX(\tau)}{d \tau} \right)_{\mbox{\tiny back \normalsize}}&=& \frac{c^2}{2}\int^\tau_{\tau_0}d\tau '[\phi_0(x(\tau )) , \phi_0(x(\tau )) ]\{M_0(\tau'),[M_0(\tau),X_0(\tau)]\}. \nonumber\\
\end{eqnarray}
To analyse the variations of the brane's energy due to the vacuum fluctuations and back reaction we consider $X(\tau)=H(\tau)$.
We now take the expectation values of these operators in the state $|0,E_i>$, where the scalar field is in its vacuum state and the brane is in an energy eigenstate of $E_i$. This provides us with the following results
\begin{eqnarray}
<0,E_i|\left ( \frac{dH(\tau)}{d \tau} \right)_{\mbox{\tiny vac \normalsize}}|0,E_i>&=& 2ic^2\int^\tau_{\tau_0}d\tau 'C_\phi(x(\tau) , x(\tau'))\frac{d}{d\tau}\chi_i (\tau, \tau') \nonumber\\
<0,E_i|\left ( \frac{dH(\tau)}{d \tau} \right)_{\mbox{\tiny back \normalsize}}|0,E_i>&=& 2ic^2\int^\tau_{\tau_0}d\tau '\chi_\phi (x(\tau ) , x(\tau'))\frac{d}{d\tau}C_i (\tau, \tau') \nonumber\\
\end{eqnarray}
where we have introduced the two-point functions \footnote{In atomic theory, the functions $C_\phi$ and $C_i$ are called the two-point correlation functions, while $\chi_\phi$ and $\chi_i$ are the linear suspectibilities.}
\begin{eqnarray}
 2C_\phi(x(\tau), x(\tau '))&=&<0|\{\phi_0(x(\tau)), \phi_0 (x(\tau'))\}|0> \nonumber\\
 2\chi_\phi(x(\tau), x(\tau '))&=&<0|[\phi_0(x(\tau)), \phi_0 (x(\tau'))]|0>\nonumber\\
 2C_i(\tau, \tau ')&=&<E_i|\{M_0(\tau), M_0(\tau')\}|E_i> \nonumber\\
 2\chi_i(\tau, \tau ')&=&<E_i|[M_0(\tau), M_0(\tau')]|E_i>.
\end{eqnarray}
For a simple model with discrete energy levels we find that
\begin{eqnarray}
C_k(\tau, \tau ')&=&\sum_j |<E_j|M_0(0)|E_k>|^2 \cos((E_j -E_k)\Delta\tau)\nonumber\\
\chi_k(\tau, \tau ')&=&-\sum_j |<E_j|M_0(0)|E_k>|^2i\sin((E_j -E_k)\Delta\tau). \nonumber
\end{eqnarray}
We may obtain the two-point functions for the scalar field via the Wightman function
\begin{eqnarray}
 2C_\phi(x,x')&=& {\cal G}^+(x,x') +{\cal G}^+(x',x)\nonumber\\
 2\chi_\phi(x,x')&=& {\cal G}^+(x,x')-{\cal G}^+(x',x).
\end{eqnarray}
It must be noted that the structure of the Wightman function in section \ref{wight} can in some sense interchange $C_\phi $ and $\chi_\phi $, depending on the dimension. This effect is linked to the different statistics we obtain in even and odd dimensions.   
\newpage
\section{The heating of the brane by quantum fluctuations}
With the formalism in place to analyse the excitations of the brane in detail, we may use the Wightman function derived in section \ref{wight} to calculate the vacuum fluctuations and back reactions that the brane experiences as it moves through the bulk. The notion of a well defined temperature is of course restricted to cosmological time scales over which the acceleration of the brane universe does not change appreciably. The vacuum fluctuations are obtained from
\begin{eqnarray}
<0,E_i|\left ( \frac{dH(\tau)}{d \tau} \right)_{\mbox{\tiny vac \normalsize}}|0,E_i>&=& 2ic^2\int^\infty_{-\infty}d\tau 'C_{\phi}(x(\tau) , x(\tau '))\frac{d}{d\tau}\chi_i (\tau, \tau'). \nonumber\\
\end{eqnarray}
For the constant acceleration trajectories we have that
\begin{eqnarray}
 C_{\phi}(\Delta \tau )&=& \frac{1}{2} \left ( {\cal G}^+(\Delta \tau -i \epsilon )+ {\cal G}^+(-\Delta \tau - i \epsilon ) \right ).
\end{eqnarray}
The calculations are similar to those of the previous section and the vacuum fluctuations for the supercritical trajectories are found to be
\begin{eqnarray}
 <0,E_i|&&\left ( \frac{dH(\tau)}{d \tau} \right)_{\mbox{\tiny vac\normalsize}}|0,E_i>=  \nonumber\\
&&-2c^2\sum_{E_{ji}<0} |M_{ji}|^2 |E_{ji}| F (\frac{|E_{ji}|}{ \omega} )\left [ 1 + \frac{(-1)^N + 1}{e^{2 \pi \frac{|E_{ji}|}{ \omega} } - (-1)^N} \right ] \nonumber \\
&& +2c^2\sum_{E_{ji}>0} |M_{ji}|^2 |E_{ji}|F (\frac{|E_{ji}|}{\omega} ) \left [ 1 + \frac{(-1)^N +1}{e^{2 \pi \frac{|E_{ji}|}{ \omega} } - (-1)^N} \right ]  \nonumber \\
\end{eqnarray}
with
\begin{eqnarray}
 E_{ji}&=& E_j -E_i\nonumber\\
M_{ji} &=& <E_j|M_0(0)|E_i>.
\end{eqnarray}
We see that vacuum fluctuations are unaffected in \textit{odd} dimensions and enhanced in \textit{even} dimensions.\\  
The back reaction is calculated similarly and we find that along the supercritical trajectories it is given by  
\begin{eqnarray}
 <0,E_i|&&\left ( \frac{dH(\tau)}{d \tau} \right)_{\mbox{\tiny back\normalsize}}|0,E_i>= \nonumber\\
&& -2c^2\sum_{E_{ji}<0} |M_{ji}|^2 |E_{ji}| F (\frac{|E_{ji}|}{ \omega} )\left [ 1 + \frac{(-1)^N - 1}{e^{2 \pi \frac{|E_{ji}|}{ \omega} } - (-1)^N} \right ] \nonumber \\
&& -2c^2\sum_{E_{ji}>0} |M_{ji}|^2 |E_{ji}|F (\frac{|E_{ji}|}{\omega} ) \left [ 1 + \frac{(-1)^N -1}{e^{2 \pi \frac{|E_{ji}|}{ \omega} } - (-1)^N} \right ].  \nonumber \\
\end{eqnarray}
The back reactions behave in a complementary way to the vacuum fluctuations: they are unaffected in \textit{even} dimensions and are modified in \textit{odd} dimensions.  
The net effect of this is that the total rate of change of the brane's energy is given by
\begin{eqnarray}\label{total}
 <0,E_i|&&\left ( \frac{dH(\tau)}{d \tau} \right)_{\mbox{\tiny total \normalsize}}|0,E_i>= \nonumber \\
&&-2c^2\sum_{E_{ji}<0} |M_{ji}|^2 |E_{ji}|F (\frac{|E_{ji}|}{\omega} ) \left [\frac{e^{2 \pi \frac{|E_{ji}|}{ \omega} }}{e^{2 \pi \frac{|E_{ji}|}{ \omega} } - (-1)^N} \right ] \nonumber \\
&& +2c^2\sum_{E_{ji}>0} |M_{ji}|^2 |E_{ji}|F (\frac{|E_{ji}|}{ \omega} ) \left [ \frac{1}{e^{2 \pi \frac{|E_{ji}|}{ \omega} } - (-1)^N} \right ]. \nonumber \\
\end{eqnarray}
It is evident from this that an accelerated system in its ground state will experience excitations.
\subsection{Characteristic Time}
It is instructive to obtain a time scale over which any heating from this coupling takes place. To achieve this we may consider the simplest case of a two level system, with energies $\pm\frac{1}{2}\Delta E$ and $M_{ij}$ being a symmetric matrix.\\
Then for the system being in a general state, equation (\ref{total}) may be written as
\begin{eqnarray}
 <\dot{H}>&=& \frac{-\frac{1}{4}c^2 \Delta E F(\frac{\Delta E}{\omega })}{e^{2\pi \frac{\Delta E}{\omega }}- (-1)^N}\left [ (e^{2 \pi \frac{\Delta E}{\omega }} -1) + ( e^{2 \pi \frac{\Delta E}{\omega }} +1) \frac{<H(\tau )>}{\frac{1}{2}\Delta E} \right ]. \nonumber   
\end{eqnarray}
The solution of which is
\begin{eqnarray}
 <H(\tau)>&=& E_e - (E_e - <H(0)>)e^{-\frac{\tau}{\tau_e}}
\end{eqnarray}
with
\begin{eqnarray}
 \tau_e&=& \frac{e^{2 \pi \frac{\Delta E}{\omega }} - (-1)^N}{\frac{1}{2}c^2 F (e^{2 \pi \frac{\Delta E}{\omega }} +1)}  \\
E_e &=& -\frac{1}{2}\Delta E \tanh(\frac{\pi \Delta E}{\omega }).    
\end{eqnarray}
We may interpret $\tau_e$ as the characteristic time for the system to evolve into thermal equilibrium with the scalar field. The energy of the system in equilibrium is then given by $E_e$ and is clearly raised above above the ground state for all accelerations and this equilibrium value is independent of whether the dimension is even or odd. We also have the simple result that
\begin{eqnarray}
 \tau_e &=&\frac{2}{c^2F} \hspace{2cm}\mbox{in odd dimensions}\nonumber\\
&& \frac{4 |E_e|}{c^2F\Delta E} \hspace{1.4cm}\mbox{in even dimensions.} 
\end{eqnarray}
\subsection{Asymptotic temperature}
We saw previously that a brane, obeying the junction conditions, accelerates. For a single symmetric brane we found that this was given by
\begin{eqnarray}
 a&=&\frac{\kappa ^ 2}{6}( 2 \rho + 3 p - \sigma  ).
\end{eqnarray}
In general this is not a constant value, but varies in time with the density and pressure. We can however consider asymptotics, where the variation of the density becomes small, and so we are justified in attaching a well-defined temperature for the bulk field. For such a universe, moving through an AdS spacetime, a bulk scalar field in its vacuum state would be perceived to be at a temperature
\begin{eqnarray}
 T&=&\frac{\kappa ^2\sqrt{\rho(\rho - \sigma) + \frac{3\Lambda_4}{\kappa ^4}}}{6 \pi }
\end{eqnarray}
where we have used the results from section 2.\\
In the case where $\sigma > \frac{2\sqrt{3\Lambda_4}}{\kappa^2}$ it must be understood that for values of $\rho $ between $\rho_1 = \frac{\sigma}{2} - \sqrt{\frac{\sigma^2}{4} -\frac{3\Lambda_4}{\kappa^4}}$ and  $\rho _2 =\frac{\sigma}{2} + \sqrt{\frac{\sigma^2}{4} -\frac{3\Lambda_4}{\kappa^4}}$ the temperature of the bulk is measured to be zero since the motion of the brane is subcritical in the bulk spacetime, see figure 1.
\begin{figure}[!ht]
\begin{center}
\epsfig{file=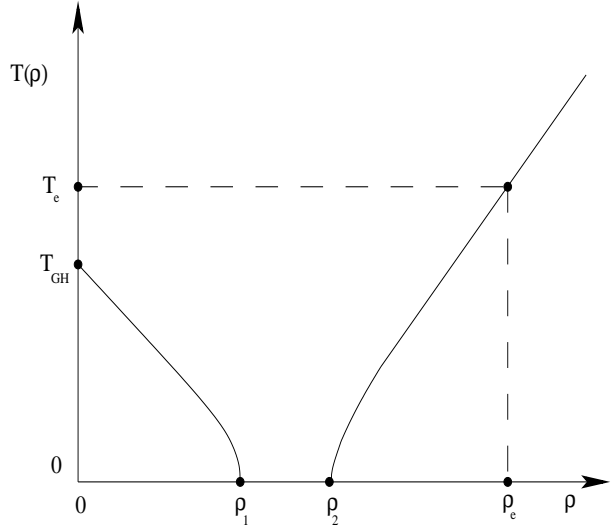 ,height=7cm,width=8cm, angle=0}
\end{center}
\caption{The temperature of a bulk AdS field in its vacuum as a function of the density of matter fields on the brane.}
\end{figure}
It is clear that in the limit $\rho \rightarrow 0$ the temperature of the bulk becomes 
\begin{eqnarray}
T(0) =\frac{1}{2 \pi }\sqrt{\frac{\Lambda_4}{3}}.
\end{eqnarray}
This is the Gibbons-Hawking temperature of a universe with cosmological constant $\Lambda _4$. This would be the asymptotic temperature of the bulk field if no energy exchange was permitted between the brane and the bulk thermal state. If energy exchange is allowed, the possibility exists that the universe will settle down to an equilibrium temperature $T_e$, which is above the zero density value. We can see that at least one other point exists where the brane is at the same temperature as the bulk since away from $\rho _1$ and $\rho _2$ the bulk temperature varies almost linearly with $\rho $, while the brane temperature is generally a fractional power of $\rho $. If the thermal coupling between the brane and bulk is strong the evolution of the temperatures is complicated. The issue of whether, for example, the current thermal state of the universe is determined from couplings to a bulk spacetime would depend on, among other things, the properties of the bulk and the heating time scale. Whether any thermal equilibrium point is stable or not will depend on the rate of the thermal exchange between the brane and bulk\cite{smolin}. If it is stable then the universe will remain at a non-zero density and temperature, $\rho_e$ and $T_e$. If this point is unstable the density of the universe will drop below $\rho_2$, where its bulk motion becomes subcritical and the Unruh temperature of the bulk vanishes. The temperature of the bulk will switch back on when the density of the universe passes below $\rho_1$, and increases until thermal equilibrium is achieved at the Gibbons-Hawking temperature $T_{GH}$. 
\section{Conclusion}
In this paper we made the observation that, in general, branes accelerate through their bulk spacetimes and so any fields, coupled to matter on the brane and in their vacuum state, will be perceived as a thermal reservoir. Effectively the matter fields of the universe play the role of an accelerated particle detector in the bulk spacetime.\\
We restricted our attentions to AdS spacetimes, and in particular to geodesics and constant acceleration trajectories. The latter trajectories being further subdivided into three cases, depending on whether the acceleration was greater than, equal to, or less than the mass scale $k$ of the spacetime.\\
We obtained a consistent set of limits on the Wightman functions for $N$-dimensional AdS spacetime and derived the Wightman function for a conformally coupled, massles scalar field in such a spacetime.\\
We showed that the transfer function vanished for all of the basic stationary trajectories except for those with a supercritical acceleration. In this case we obtain a thermal distribution which is fermionic in odd dimensions, bosonic in even dimensions, and possessing a polynomial weighting which was curvature dependent. We analysed these thermal excitations in more detail by calculating the vacuum fluctuations and back reactions seperately for a discrete, multilevel system coupled to the scalar field. For this we were able to extract a characteristic time for the process. \\
We then considered the asymptotic behaviour of a brane universe, whose acceleration through the bulk approaches a contant value. For a non-empty brane, thermal equilibrium with the bulk can occur at a non-zero density.  It was also shown that if the density of the brane is zero then any fields in their vacuum states would be perceived to be at the Gibbons-Hawking temperature $T_{GH}=\frac{1}{2 \pi}\sqrt{\frac{\Lambda _4}{3}} $. In this context it is possible to recast the Gibbons-Hawking effect as an Unruh effect in an embedding spacetime.
\section*{Acknowledgements}
It is a pleasure to thank Prof. Anne Davis and Prof. Ian Drummond for useful and friendly conversations. 
This work is funded in part by EPSRC and by St. John's College, Cambridge.
\newpage
\section{Appendix}
Here we gather some results that were used in the paper.
\subsection{Polynomial in $\frac{E }{\omega} $}
Given
\begin{eqnarray}
 G(z)&=& \frac{e^{-iEz}}{\sinh^{N-2} (\frac{\omega z}{2} )}
\end{eqnarray}
we calculate its residues at the poles $z_m = -\frac{2m \pi i  }{\omega} $ via
\begin{eqnarray}
 \mbox{Res}(G(z))|_{z_m}&=&\lim_{z \rightarrow z_m} \frac{1}{(N-3)!}\frac{d^{N-3}}{dz^{N-3}}\left [ (z-z_m)^{N-2} f(z) \right ]
\end{eqnarray}
which we find to be of the form
\begin{eqnarray}
 \mbox{Res}(G(z))|_{z_m}&=&\frac{if(\frac{E}{ \omega} )}{(2 \pi)\omega } \left ( (-1)^N e^{-2 \pi \frac{E }{\omega} } \right )^m.
\end{eqnarray}
The explicit forms of the polynomials for dimensions $N=3,4, \cdots 10$, are given as
\begin{eqnarray}
f(x)&=&4 \pi , \hspace{0.3cm}(N=3) \nonumber \\ 
 &=&8 \pi x, \hspace{0.3cm}(N=4) \nonumber \\ 
 &=&2\pi  (1+4x^2), \hspace{0.3cm}(N=5) \nonumber \\ 
 &=&\frac{16 \pi }{3}(x+x^3), \hspace{0.3cm}(N=6) \nonumber \\ 
 &=&2 \pi (\frac{3}{4} + \frac{10}{3}x^2 +\frac{4}{3}x^4),\hspace{0.3cm} (N=7) \nonumber \\ 
 &=&\frac{16 \pi }{15}(4x +5x^3 + x^5),\hspace{0.3cm} (N=8) \nonumber \\ 
 &=&2\pi ( \frac{5}{8} + \frac{259}{90}x^2 +\frac{14}{19}x^4 + \frac{8}{9}x^6),\hspace{0.3cm} (N=9) \nonumber \\ 
 &=&\frac{32 \pi }{315}(36x +48x^3 + 14x^5 + x^7), \hspace{0.3cm}(N=10) \nonumber \\ 
\end{eqnarray}
\newpage
\subsection{The functions $q_1$ and $q_2$}
We consider the function
\begin{eqnarray}
 G(z)&=&\frac{e^{-iEz} }{\sinh^{\frac{N}{2}-1} (A + \frac{\omega z}{2})\sinh^{\frac{N}{2}-1} (A - \frac{\omega z}{2})}
\end{eqnarray}
which has poles at the points $z_{m,\pm} = \pm \frac{2A}{\omega } - \frac{2m \pi i }{\omega} $.\\
For $N$ even we may calculate the residues at these poles by the standard method to find that
\begin{eqnarray}
 \mbox{Res}(G(z))|_{z_{m,+}} &+&\mbox{Res}(G(z))|_{z_{m,-}} = \nonumber\\
&&\frac{ie^{-2m\pi \frac{E}{\omega} }}{2 \pi} \left (\frac{k}{\omega}\right)^{N-2}( q_1(E,a,k) \sin \theta + q_2(E,a,k) \cos \theta ).\nonumber
\end{eqnarray}
The explicit form of these functions for $N =4,6,8,10$ are given by
\begin{eqnarray}
 q_1&=&\frac{4\pi}{a}, q_2=0,\hspace{0.4cm} (N=4)\nonumber\\
q_1&=& 2\pi( \frac{2}{a} -\frac{k^2}{a^3}), q_2 =2\pi( \frac{-2E}{a^2}),\hspace{0.4cm} (N =6) \nonumber\\
q_1 &=&2\pi(\frac{3k^2}{4a^5} -\frac{2k^2}{a^3} -\frac{E^2}{a^3} +\frac{2}{a}), q_2 =2\pi( -\frac{3Ek^2}{a^4} - \frac{3E}{a^2}),\hspace{0.4cm} (N =8) \nonumber \\
q_1&=& 2\pi(\frac{9k^4}{4a^5} - \frac{5k^6}{8a^7} +\frac{2E^2}{a^3} + \frac{3k^2}{a^3}+\frac{k^2E^2}{a^5} +\frac{2}{a}),\hspace{0.4cm} (N=10)\nonumber \\
q_2 &=& 2\pi (\frac{E^3}{3a^4} -\frac{11E}{3a^2}+\frac{11Ek^2}{3a^4}-\frac{5k^4}{4a^6}),\hspace{0.4cm} (N=10)\nonumber
\end{eqnarray}
When $N$ is odd, we have branch points to deal with and the problem cannot be tackled by residues.\\ We may nevertheless deduce the thermal distribution from the antiperiodicity of the Wightman function:
\begin{eqnarray}
 {\cal G}^+_2( \Delta \tau + \frac{2 \pi i}{\omega })&=& (-1)^{N} {\cal G}^+_2( \Delta \tau). 
\end{eqnarray}

\end{document}